\documentclass[12pt,preprint]{aastex}
\usepackage{float}
\usepackage{graphicx}
\usepackage{epsfig}
\usepackage[perpage,symbol*]{footmisc}
\usepackage{natbib}
\usepackage{dcolumn}
\usepackage{color}
\usepackage[figureright]{rotating}
\usepackage{amsmath} 
\usepackage{CJK}
\usepackage{subfigure}

\def\apjl{ApJ}
\def\apjs{ApJ}

\def\aap{A\&A}

\shorttitle{Magnetar Simulation}
\shortauthors{G.R. Yao et al.}

\begin{document}
\begin{CJK}{GB}{gbsn}

\title{MAGNETAR GIANT FLARES IN MULTIPOLAR MAGNETIC FIELDS. III. \\
MULTIPOLAR MAGNETIC FIELD STRUCTURE VARIATIONS}


\author{Guang-Rui~Yao (姚广瑞)\altaffilmark{1,2}, Lei~Huang (黄磊)\altaffilmark{1,3}, Cong~Yu (余聪)\altaffilmark{4}, Zhi-Qiang~Shen (沈志强)\altaffilmark{1,3}}

\altaffiltext{1}{Shanghai Astronomical Observatory, Chinese Academy of Sciences, Shanghai 200030, China}
\altaffiltext{2}{University of Chinese Academy of Sciences, Beijing 100049, China}
\altaffiltext{3}{Key Laboratory of Radio Astronomy, Chinese Academy of Sciences, Nanjing 210008, China}
\altaffiltext{4}{School of Physics and Astronomy, Sun Yat-Sen University, Guangzhou, 519082, China
}


\begin{abstract}

We have analyzed the multipolar magnetic field structure variation at neutron star surface by means of the catastrophic eruption model, and find that the variation of the geometry of multipolar fields on the magnetar surface could result in the catastrophic rearrangement of the magnetosphere, which provides certain physical mechanism for the outburst of giant flares. The magnetospheric model we adopted consists of two assumptions: a helically twisted flux rope is suspended in an ideal force-free magnetosphere around the magnetar, and a current sheet emerges during the flux rope evolution. Magnetic energy accumulates during the flux rope's gradual evolution along with the variation of magnetar surface magnetic structure before the eruption. The two typical behaviors, either state transition or catastrophic escape, would take place once the flux rope loses equilibrium, thus tremendous accumulated energy is radiated. We have investigated the equilibrium state of the flux rope and the energy release affected by different multipolar structures, and find structures that could trigger violent eruption and provide the radiation approximately 0.5$\%$ of the total magnetic energy during the giant flare outburst. Our results provide certain multipolar structures of the neutron star's magnetic field with an energy release percentage 0.42$\%$ in the state transition and 0.51$\%$ in the catastrophic escape case, which are sufficient for the previously reported energy release from SGR 1806-20 giant flares.

\end{abstract}




\newpage
\section{INTRODUCTION}\label{sec:intro}

Magnetars are essentially a type of neutron stars surrounded by extremely strong ($10^{14}-10^{15}$ G) magnetic fields; and there are two independently discovered kinds of magentars: soft gamma-ray repeaters (SGRs) and anomalous X-ray pulsars (AXPs) \citep{1979Natur.282..587M, 1992ApJ...392L...9D, 2002Natur.419..142G}. Unlike the common rotation-powered neutron stars, it is commonly believed that the magnetic energy dissipation is the ultimate source for their high energy persistent radiation and transient bursting emission \citep{2006csxs.book..547W, 2008A&ARv..15..225M}. One of the most extraordinary phenomena correlated with magnetars is the particular spasmodic giant flare, which is characterized by the sudden release of tremendous amounts of energy.

Until now, there have been only three detected giant flare events, all of which are observed from SGRs; respectively: the 1979 flare from SGR 0526-66 \citep{1982Ap&SS..84..173M}, the 1998 flare from SGR 1900+14 \citep{1999ApJ...510L.115K}, and the 2004 flare from SGR 1806-20 \citep{2005ApJ...624L.105M}. A large amount of energy is released instantaneously during these giant flare outbursts, which is about $(2-500)\times10^{44}$ ergs in a duration of a few tenths of second \citep{2008A&ARv..15..225M}. The peak luminosity during giant flare events could arrive at more than a million times of the Eddington luminosity of a normal neutron star. The peak luminosities of the 1979 flare and the 1998 flare are of the order of $10^{44}$ erg s$^{-1}$ \citep{2008A&ARv..15..225M}, however, the most extreme giant flare was observed from SGR 1806-20 in 2004, with a peak luminosity approximate to $2\times10^{47}$ erg s$^{-1}$ and an isotropic energy release of about $(1.8-4.6)\times10^{46}$ erg \citep{2005Natur.434.1098H, 2005Natur.434.1110T}. On the other hand, the short timescale of the giant flare rise time, $\sim$0.25ms \citep{2005Natur.434.1107P}, would greatly restrict the theoretical models to interpret these bursting.

Giant flares demonstrate common properties in the X-rays; specifically, the light curves consist of an initial hard spike (maintaining a fraction of a second) followed by a longer pulsating tail (maintaining several minutes) \citep{2005Natur.434.1110T, 2005ApJ...624L.105M}. During the initial spike, the luminosities of giant flares increase rapidly as a fraction of the energy escapes directly as a relativistic expanding plasma. However, the following pulsating tail remains confined in a trapped thermal fireball near the neutron star surface restricted by the strong magnetic field \citep{1995MNRAS.275..255T}; the remaining energy is gradually radiated during this process. The pulse profile of the pulsating tail of the 1998 giant flare from SGR 1900 +14 showed a four-peak pattern, which was thought to be direct evidence of the multipolar structure of the magnetar magnetic field \citep{2001ApJ...549.1021F}. In addition, the multi-peak structure during the 2004 giant flare from SGR 1806-20 might suggest that the flux distribution on the surface involves higher order multipole\citep{2011ApJ...729....1X}. Further more, \citet{2011ApJ...729....1X} found two phases drifting towards opposite directions after fitting the sub-pulses after giant flare during the pulsating tail, which might be another evidence for the existence of multipolar structures. 

The instantaneously released enormous energy, as a typical characteristic of giant flares mentioned above, could be well explained by a magnetospheric model put forward by \citet{2006MNRAS.367.1594L}. The author holds the opinion that the tremendous energy released during the giant flare outburst accumulated gradually in the magnetar magnetosphere before the eruption, and provides certain physical mechanism to interpret the transient rise time of the giant flare. Based on the magnetospheric model assumption, a catastrophic flux rope eruption model has been proposed to demonstrate the magnetic field energy storage process \citep{2012ApJ...757...67Y}. Yu describes the mechanism for driving an eruption by the quasi-static evolution of a helically twisted flux rope in the magnetosphere. Then the current sheet hypothesis has been put forward by \citet{2013ApJ...771L..46Y} under the circumstance of a dipolar magnetic field. Further more, in order to explain the existence of multipoles observed during an giant flare event \citep{2001ApJ...549.1021F, 2011ApJ...729....1X}, multipolar magnetic boundary variations at the magnetar surface have been taken into consideration. Therefore, in \citet{2014ApJ...787..175H}, hereafter Paper I, the energetics of the flux rope eruption model in the multipolar field background are carefully investigated. It is found that the accumulated magnetic energy before eruption in this model could be sufficient to drive a giant flare. Subsequently, in \citet{2014ApJ...796....3H}, hereafter Paper II, the effects of the current sheet in multipolar background are taken into consideration, which bring new features contrasting to the previous calculation. They mainly focus on how the magnetar surface magnetic flux gradually alters, which is assumed to be increased by the injections from the interior of magnetar, that could trigger the outburst and account for the enormous energy release. However, the multipolar structure variation (represents the crust gradual motions in actual scenarios), as another important potential origin for outburst, is not taken into consideration in Paper II and requires further investigation.

We have investigated the distinctive multipolar structure variation with the help of the flux rope eruption model. The evolution of the flux rope could represent the energy accumulation process before the eruption and the sudden outburst of the giant flare. We finally find certain circumstances that magnetic multipolar structure variation could lead to the state transition or catastrophic escape of the flux rope and might account for the enormous energy release during the outburst of the known giant flares.

\section{ERUPTION MODEL}

In the magnetar magnetosphere, factors like the pressure and inertia of the plasma could be ignored because of the dominating strong magnetic fields \citep{2002ApJ...574..332T, 2011MNRAS.411.2461Y}. Therefore, an ideal force-free magnetic field is assumed to simulate the magnetosphere around the magnetar, in which \textbf{J}$\times$ \textbf{B}=0. The inhomogeneous Grad-Shafranov (GS) equation \citep{2002ApJ...574..332T} is adopted to determine the axisymmetric magnetic field configurations.

\subsection{Axisymmetric Force-free Magnetosphere with Helically Twisted Flux Rope and Current Sheet}

In our magnetospheric model, we adopt a toroidal and helically twisted flux rope as a perfect probe to demonstrate the global physical mechanism in the magnetosphere before eruption. The giant flare precursor might be correlated to the evolution of the helically twisted flux ropes \citep{2010MNRAS.407.1926G}. Then the magnetic field and current density distribution inside the flux rope are obtained by means of the force-free solution \citep{1950Ark...2..361}. To be specific, the neutron star surface radius is $r_s$; the twisted flux rope has a major radius, $h$, which could be comprehended as the height of the flux rope (measured from the magnetar center) and a minor radius, $r_0$, which is micro compared to $h$. The flux rope separates the magnetosphere into two parts: the region inside the flux rope, and the region outside the flux rope. 

The magnetic fields $\mathbf{B}$ in the magnetosphere (outside the flux rope) is assumed to be axisymmetric and is described in spherical coordinates ($r$, $\theta$, $\phi$),
\begin{equation}
\mathbf{B}=-\frac1{r^2}\frac{\partial\Psi}{\partial\mu}\hat{\mathbf{e}}_r-\frac1{r\sin\theta}\frac{\partial\Psi}{\partial r}\hat{\mathbf{e}}_\theta,
\end{equation}
in which $\mu=\cos\theta$, and $\Psi=\Psi(r,\mu)$ is the magnetic steam function. The force-free condition could be expressed in terms of the standard GS equation:
\begin{equation}
\frac{\partial^2\Psi}{\partial r^2}+\frac{1-\mu^2}{r^2}\frac{\partial^2\Psi}{\partial \mu^2}=-r\sin\theta\frac{4\pi}cJ_\phi,
\end{equation}
where $c$ represents the velocity of light. $J_\phi$ is the current density in the form \citep{2000Pri...}
\begin{equation}
J_\phi=\frac{I}{h}\delta(\mu)\delta(r-h),
\end{equation}
where $I$ represents the electric current inside the flux rope and $\delta$ is the Dirac $\delta$ function. In order to calculate conveniently, a dimensionless current $J$ is adopted in the following calculation, where $J\equiv I/I_0$, $I_0={r_0I}/{r_{00}}$, and we take $r_{00}=0.01$ \citep{2012ApJ...757...67Y}.

The boundary condition at the hypothetical spherical surface of magnetar ($r=r_s$) is:
\begin{equation}
\Psi_s(r_s,\mu)=\Psi_0\sigma\Theta(\mu),
\end{equation}
where $\Psi_0$ is a magnetic flux constant parameter, the dimensionless quantity $\sigma$ stands for the magnetic flux magnitude at the neutron star surface, and $\Theta(\mu)$ represents the separated variable of the stream function - the angular dependent component.\newline 

Eventually, we give a brief description of another characteristic assumption in our model, the current sheet, which appears in the magnetosphere during the eruption. The current sheet lies at the equatorial plane where $\theta=\pi/2$, and ranges from the magnetar surface (r=$r_s$, $\theta=\pi/2$) to the endpoint (r=$r_1$, $\theta=\pi/2$). The existence of the current sheet requires the following boundary condition to be satisfied as Paper II,
\begin{equation}
\Psi(r,0)\equiv\Psi_{cs} \quad  (r_s\leqslant r\leqslant r_1),
\end{equation}
where $\Psi_{cs}$ stands for the constant value along the current sheet, $\Psi_{cs}=\Psi_0\sigma\Theta_0$ for $\Theta_0=\Theta(0)$.

\subsection{Multipolar Boundary Conditions and the Equilibrium Equations}

We assume that a magnetar giant flare is primarily driven by the catastrophic instability caused by the loss of force equilibrium in the magnetospheric system, which is initiated by boundary conditions or changes at the magnetar surface. In the multipolar background field, the equilibrium curve of the flux rope has been carefully investigated in Paper II. The loss of equilibrium triggers the state transition directly during the evolution before eruption. Then, another stable equilibrium has been established soon after the transition. Thus along the stable branch of the equilibrium curve, the succedent evolution of the flux rope maintains in a quasi-static process. The observed multi-peak profile during the 2004 giant from SGR 1806-20 suggests that the flux distribution on the magnetar surface involves higher order multipole \citep{2011ApJ...729....1X}. So in order to solve the GS equation (2), we adopt the boundary conditions including both a dipolar component and a high order multipolar component, to represent the complicated magnetic structure near the magnetar surface.

To model the multipolar field neutron star surface, we assume that the function $\Theta(\mu)$ is constituted by a dipolar component with the addition of a high order multipolar component, i.e.,
\begin{equation}
\Theta(\mu)=(1-\mu^2)+a_1 exp\left[-\frac{(\mu-\mu_0)^2}{2w^2}\right]+a_1 exp\left[-\frac{(\mu+\mu_0)^2}{2w^2}\right],
\end{equation}
where the $(1-\mu^2)$ term represents the dipolar part of the magnetic field and the remaining terms are two Gaussian functions expressing the component from multipolar structure. While the parameter $a_1$ determines the strength of the multipoles, and $\mu_0$ and $w$ describe the distribution of magnetic flux at the surface of magnetar. To be specific, the parameter $\mu_0$ (ranges between 0 and 1) determines the distance between the two supposed multipolar components at the magnetar surface: when $\mu_0$=0, the two multipolar components overlap on the equator; when $\mu_0$=1, the two components locate on the opposite poles. Therefore the varying $\mu_0$ represents the multipolar structure variation caused by the neutron crust motions. We fix $w^2$=0.001 throughout the whole calculation, thus different multipolar structures are concisely expressed. The global configuration of the magnetospheric field could be calculated numerically as Paper II. When the absolute value of the parameter $a_1$ is relatively small, the background magnetic field which is close to the equator is approximately a dipole field. The difference between dipolar-dominated field and multipolar-dominated field could be distinguished by the amount of the extreme points in the boundary flux distribution profile, which is demonstrated detailedly in Paper I. The function $\Theta(\mu)$ is symmetric to the equator ($\theta=\pi/2$), leading to a symmetric boundary condition in Eq. (4). It is clear from the distribution of $\Theta$ that as the parameter $\mu_0$ increases, the active multipolar regions which are represented by two Gaussian functions move apart from each other \citep{2012ApJ...757...67Y}.

The flux rope is assumed to be in a quasi-static equilibrium state, on a longer timescale than the dynamical flare timescale before the eruption. In order to keep the flux rope in quasi-static equilibrium states \citep{2012ApJ...757...67Y}, the following external equilibrium conditions should be satisfied. Specifically,

(a) the force balance condition, that the magnetic field generated by the electromagnetic induction of the current inside the flux rope, $B_s$ (produces the outward force), should be balanced by the magnetic field outside the flux rope, $B_e$ (produces the inward force) in the magnetosphere around the neutron star;

(b) the ideal frozen-flux condition, which requires that marginal magnetic flux stream value $\Psi(h-r_0,0)$ of the flux rope, maintains invariant during the evolution of the system before the outburst.

In order to test the multipolar structure variation, we adjust an invariable value of the magnetic flux on magnetar surface $\sigma$ in Eq. (4). Thus the two equilibrium constraints are expressed in the following two equations,
\begin{equation}
\begin{cases}
f(\mu_0,J,h)\equiv B_s-B_e=0 \\
g(\mu_0,J,h)\equiv \Psi(h-r_0,0)=\mbox{const}.
\end{cases}
\end{equation}

Given the distance between the two supposed multipolar components $\mu_0$ and a rational constant to satisfy $g(\mu_0,J,h)=\mbox{const}$, we could obtain the current $J$ and the height $h$ by adopting the Newton-Raphson method \citep{2007Pre...}. Subsequently, the variation of $h$ corresponding to $\mu_0$ that satisfies the equilibrium equation are obtained.

\section{CATASTROPHIC RESPONSE OF FLUX ROPE TO MULTIPOLAR MAGNETIC FIELD STRUCTURE VARIATIONS}

The flux rope evolution process in the magnetosphere could be demonstrated by the equilibrium curve, which is obtained after numerical calculation satisfying the equilibrium constraints at certain boundary conditions. As a result, we find two patterns, specifically, the state transition process and the catastrophic escape process. We assume that the changing $\mu_0$ of the boundary condition represents the gradual variation of the multiploar structure on magnetar surface.

\subsection{State Transition Case with the Boundary Condition Variation}

One typical example of the state transition process is demonstrated in Figure 1, in which the curve connected by three branches shows the variation of the height of flux rope $h$ with the separation between the two Gaussian structures $\mu_0$ at the surface of the magnetar. This equilibrium curve consists of three branches: stable Branch I below the point (c), unstable Branch II between the point (c) and (d), and stable Branch III above the point (d). The flux rope behaves as a harmonic oscillator along the stable branches \citep{2010hssr.book..159F, 2012ApJ...757...67Y}; on the contrary, in the unstable branch, the flux rope would deviate from the equilibrium state if any slight disturbance takes place \citep{2012ApJ...757...67Y, 2014ApJ...796....3H}. In this case, $a_1$=1.0, $\sigma$=15.47, the equilibrium curve reaches the critical point (c) of $h_c$=1.50 while $\mu_0$=0.406. Once it reached the critical point, the stable equilibrium state would no longer be maintained, thus the eruption might occur instantaneously. However, in this situation, an upper stable point (f) locates at the stable Branch III exists, where the height is $h_f$=6.49. When flux rope transits from the critical point (c) to the stable point (f), it would release enormous energy. In this case, the current sheet begins to appear near the critical point, the endpoint of the current sheet $r_1$ is described in Section 2.1. The evolution of $r_1$ is demonstrated by the dotted curve. The height of flux rope and location of the current sheet could explicitly constrain the magnetic field structure during the variations. 

The configuration of the magnetic field for the magnetar magnetosphere containing a flux rope and a current sheet under the multipolar boundary condition could be obtained with solutions of the inhomogeneous GS equation satisfying the boundary conditions and the equilibrium constraints in Section 2. Figure 2 provides the configuration of the magnetic field at the critical point (c) and the stable point (f), respectively. The inner thick solid semicircle denotes the magnetar surface and the dashed semicircle represents the position where the flux rope locates in the magnetosphere. In our simplified model, the flux rope is assumed to be a closed current ring suspended in the magnetosphere, and two ends of the flux rope are not anchored to the magnetar surface. This configuration demonstrates the quasi-static equilibrium state before the catastrophic outburst, during which the flux rope will evolve along the current sheet gradually. For the state transition case, we show two particular configurations of the magnetic field lines in Figure 2. In the left panel, the system reaches a pre-eruption state, and the current sheet starts to appear near the critical point (c). Then the flux rope is forced to move outward along the gradually increasing current sheet until it reaches a stable state. As shown in the right panel of Figure 2, the current sheet lies on the equatorial plane: the lower tip of the current is connected to the magnetar surface, while the upper tip is located at $r_1$. The right panel shows the configuration of the magnetic field when the flux rope reaches a stable state, point (f) in Figure 1. Finally, the flux rope achieves to transit from a lower unstable position to a higher stable position in the magnetosphere, during which huge energy is released.

\subsection{Catastrophic Escape Case with the Boundary Condition Variation}

The equilibrium curve of the flux rope depends on the background magnetic field. However, there exists such situation that the flux rope could not arrive at a new stable equilibrium state once the flux rope reaches a critical point. As a result, the flux rope might escape to infinity. This behavior is called catastrophic escape, which is definitely different from the normal transition process mentioned above. The eruption of the flux rope takes place on a dynamical timescale during this catastrophic escape process, when enormous amounts of magnetic energy is radiated. We provides a certain catastrophic escape case in Figure 3, where $a_1$=1.1 and $\sigma$=15.62, and the equilibrium curve is consists of three branches, which is similar to the state transition case above. As the multipolar structure changes with $\mu_0$, the equilibrium height gradually increases and finally reaches the critical point (c) with $h_c$=1.51 and $\mu_0$=0.406. Then the equilibrium curve moves along the unstable branch (Branch II) and the stable branch (Branch III). However, in this situation, Branch III extends to infinity around $\mu_0$=0.12, indicating that the flux rope will be forced to escape from the magnetosphere.

We think that either the flux rope's state transition or escaping to the infinity could be associated with the observed magnetar giant flares. The neutron star surface magnetic flux dimension could definitely affect the flux rope equilibrium state \citep{2012ApJ...757...67Y, 2014ApJ...796....3H}. However, our work mainly investigates the surface magnetic field structure, as contrast to the previous studies. We can figure out that different multipolar structures, as the boundary condition, would definitely affect the physical conditions on which the dynamical eruptions finally take place. 

In the following, we will investigate the energey carried out during the eruption of flux rope and finally find a certain background multipolar structure that can satisfy the observed catastrophic behavior.

\begin{figure}[h]
\centering
\includegraphics[width=15cm]{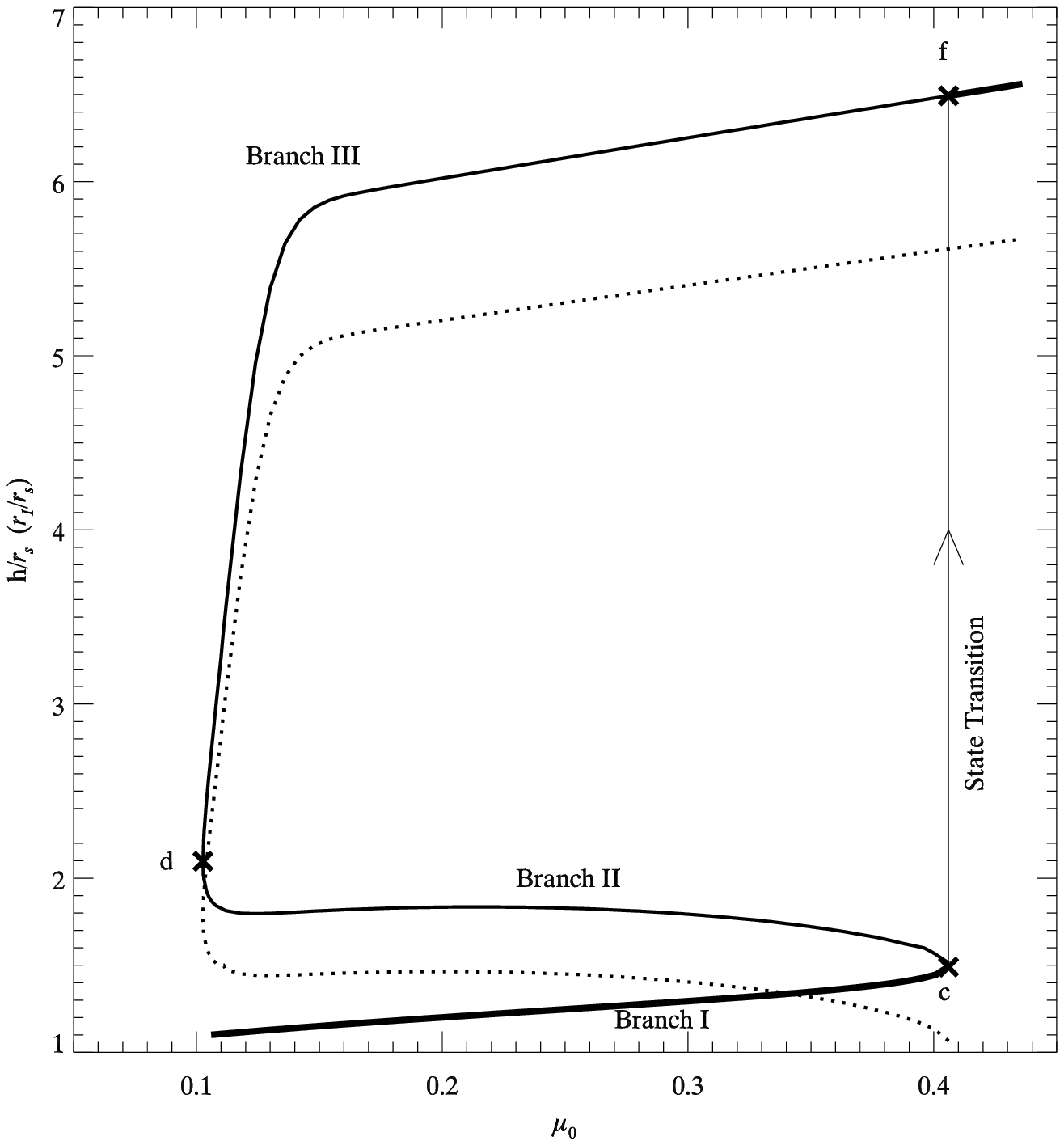}
\caption{Equilibrium height $h$ as a function of $\mu_0$ for the state transition case and the corresponding $r_1$. $a_1$=1.0 and the independent variable $\sigma$=15.47 is assumed. As $\mu_0$ increases, the equilibrium height $h$ gradually increases (Branch I) and finally reaches the critical point (c), where there is only one solution to the equilibrium equations and the equilibrium curve reaches to the unstable branch (Branch II). The critical point is approximately at $h_c$=1.50 and $\mu_0$=0.406. Then the equilibrium curve comes to the link point (d) which connects the unstable branch (Branch II)and the stable branch (Branch III), and finally jumps to the stable point (f) along the stable branch. The dotted curve shows the corresponding variation of current sheet.}
\end{figure}

\begin{figure}[htbp]
\centering 
\subfigure{ 
\begin{minipage}{7cm}
\centering 
\includegraphics[width=8cm]{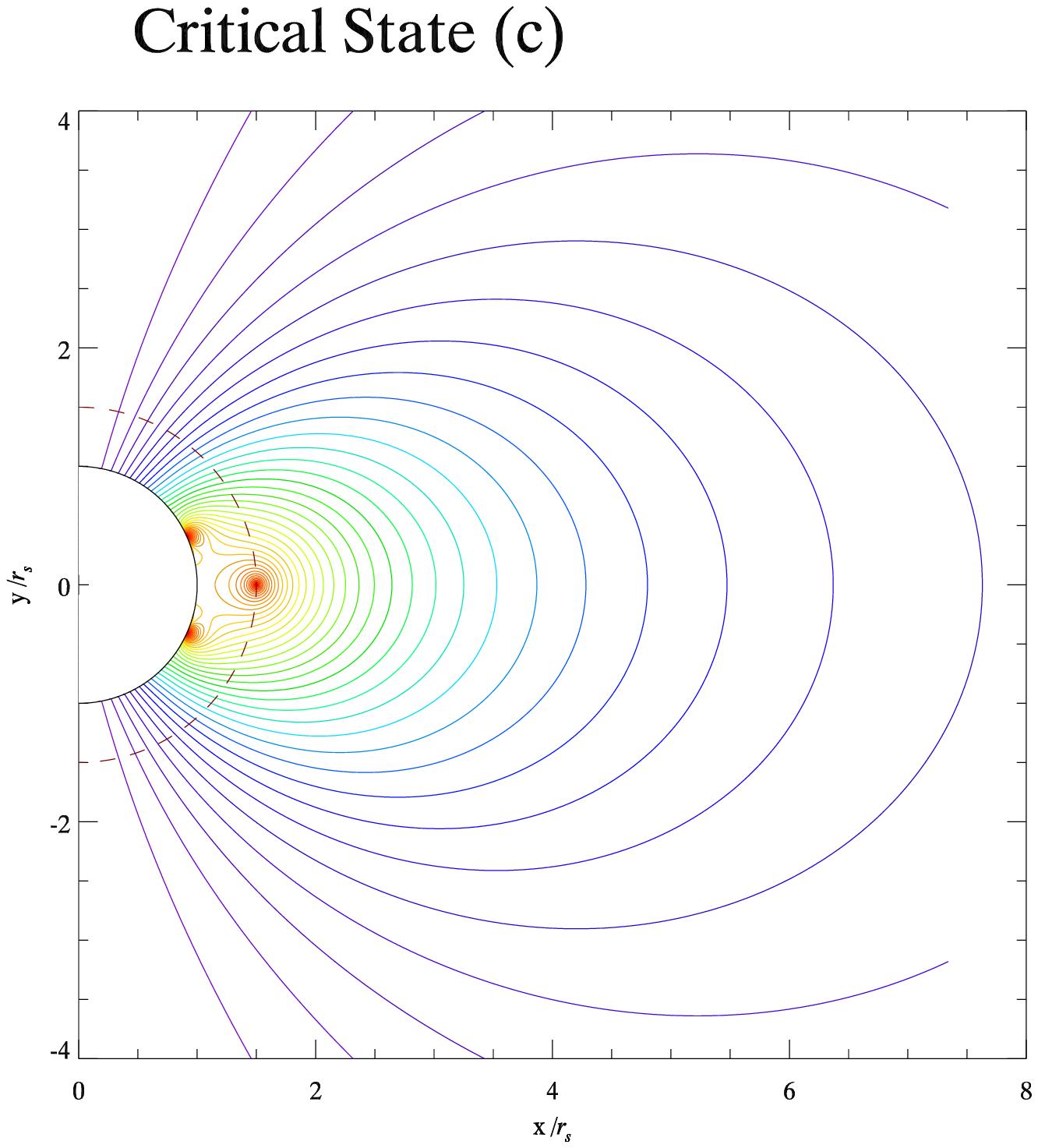}
\end{minipage}
}
\subfigure{ 

\begin{minipage}{7cm}
\centering 
\includegraphics[width=8cm]{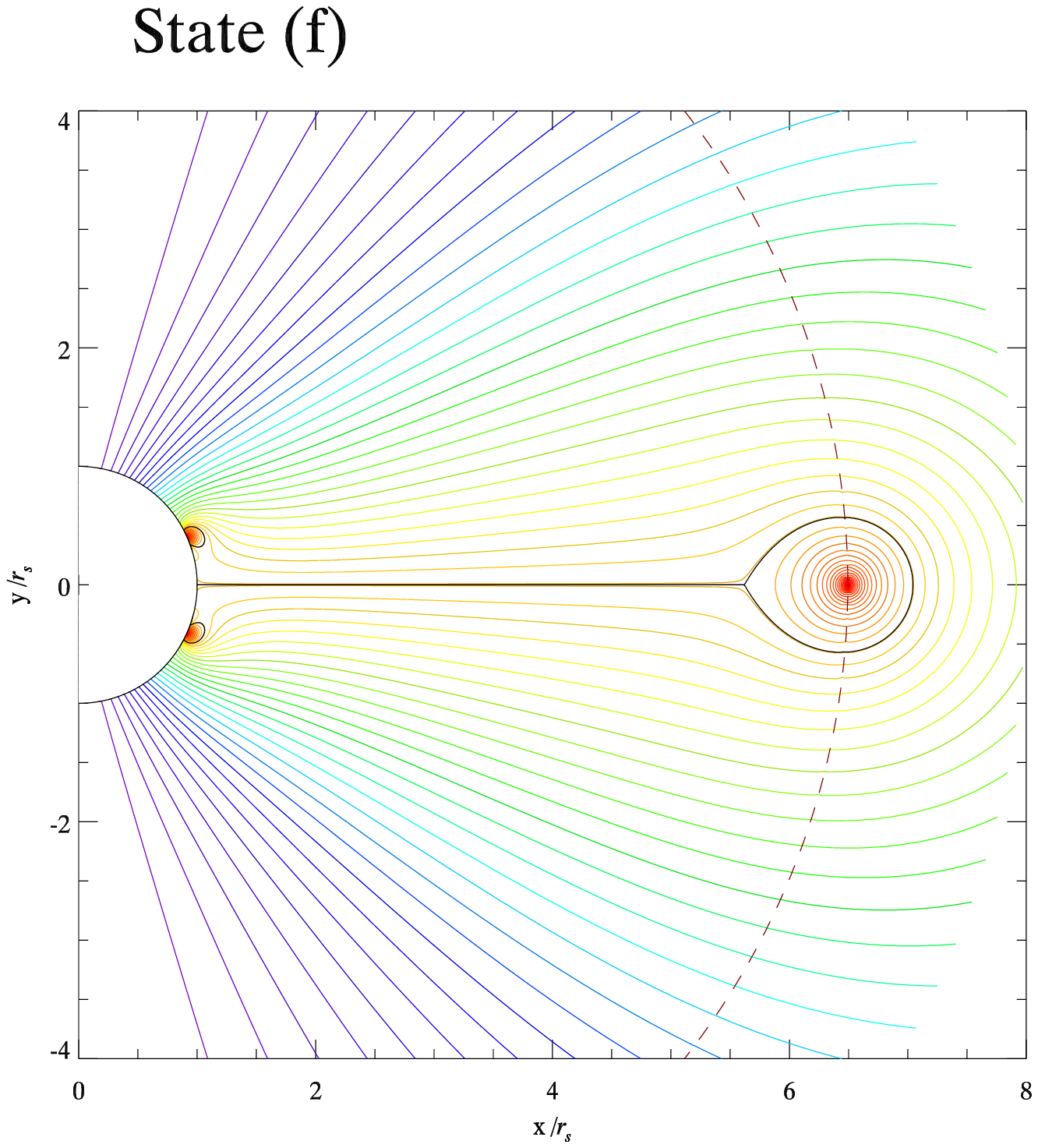}
\end{minipage}
}

\caption{Configuration of the magnetic field lines for the state transition case in Figure 1. Left: configuration at the critical point (c). The flux rope is embedded in the magnetosphere and in the pre-eruptive critical state (where $h_c$=1.50), when the flux rope is about to transit and the current sheet begins to appear. Right: configuration at the stable point (f). The flux rope is forced to move outward along the current sheet from the critical (c) to the stable point (f).} 
\end{figure}

\begin{figure}[h]
\centering
\includegraphics[width=15cm]{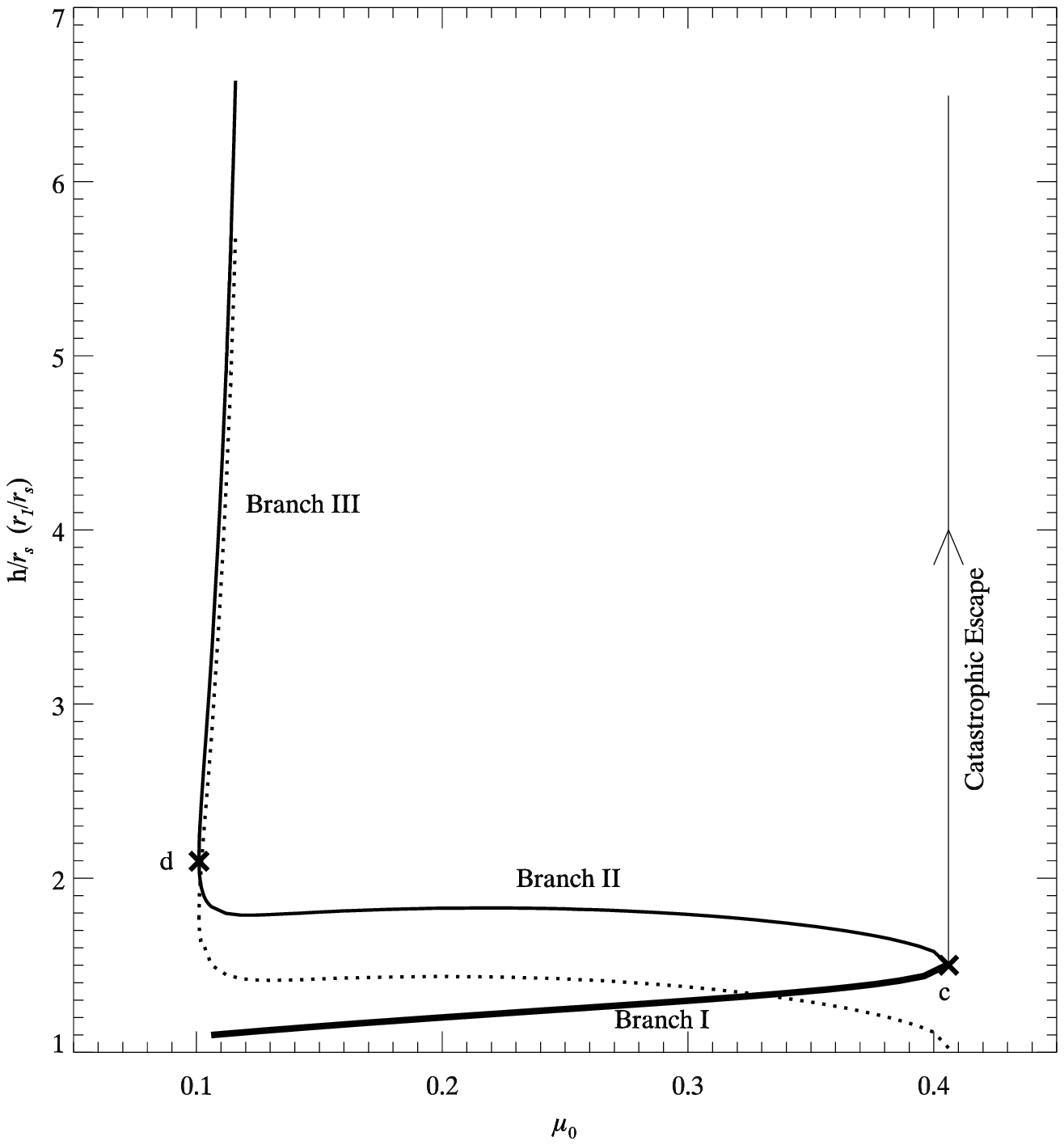}
\caption{Equilibrium height $h$ as a function of $\mu_0$ for the catastrophic escape case and the corresponding $r_1$. Here, $a_1=1.1$ and $\sigma$=15.62. As $\mu_0$ increases, the equilibrium height gradually increases and finally reaches the critical point (c), where $h_c$=1.51 and $\mu_0$=0.406. Then the equilibrium curve moves along the unstable branch (Branch II) and the stable branch (Branch III), which means the flux ropt will escape to infinity.}
\end{figure}

\subsection{Energy Release Related to Different Magnetar Surface Structures}

We further investigate the equilibrium state of the flux rope and the total energy released responding to different multipolar structures at the magnetar surface.
With the assumption of multipolar structure surface boundary conditions, we find that certain parameters in our model could generate enough energy for the reported giant flare outburst.

Once the flux rope evolves to the critical point, it will become unstable and might initiate the eruption. During the unstable process, a current sheet might appear, and the flux rope would either escape to infinity or transfer to another stable state along the current sheet. The tremendous magnetic energy stored before eruption would then be released within a transient timescale instantly.

The total magnetic energy of the magnetosphere can be calculated specifically as,
\begin{equation}
W_t(h)=W_{\rm p}-\int_{\infty}^{h}F(h')\mathrm{d}h',
\end{equation}
where the first term is the total magnetic potential energy stored in the multipolar field, and the second term represents the energy required to force the flux rope to move from infinity to the position in the magnetosphere. $F = 2\pi Ih(B_s-B_e)/c$, which is the resultant force on the idealized flux rope.
\begin{equation}
B_s=\frac{I}{ch}\left(\ln\frac{8h}{r_0}-1\right),
\end{equation}
and $B_e$ is described by \citet{2012ApJ...757...67Y}.\newline

According to magnetic virial theorem, we have\newline

\begin{equation}
W_{\rm p}=\int\frac{\mathbf{B}_{\rm p}^2}{8\pi}\mathrm{d}V=\int_{\partial V}\frac{B_{\rm p}^2}{8\pi}(\mathbf{r}\cdot\mathrm{d}\mathbf{S})-\frac1{4\pi}\int_{\partial V}(\mathbf{B}_{\rm p}\cdot\mathbf{r})(\mathbf{B}_{\rm p}\cdot\mathrm{d}\mathbf{S}),
\end{equation}
where $\mathbf{B}_{\rm p}$ is the potential magnetic field intensity, which could be obtained by calculating the partial derivative of the potential stream function $\Psi_{\rm p}$ as Eq. (1); and $\Psi_{\rm p}$ is calculated numerically by means of Legendre polynomials \citep{2012ApJ...757...67Y}.

The total magnetic energy of magnetar is approximately $E_{mag}\sim2\times10^{49} $(B/$10^{16}$ G)$^2$ erg \citep{2005Natur.434.1098H}, and the internal magnetic field  intensity B is estimated at about $(5-10)\times10^{15}$ G \citep{1996ApJ...473..322T}, which indicates that magnetic energy of magnetar is $E_{mag}\sim(5-20)\times 10^{48}$ erg. The energy release during the most powerful giant flare from SGR 1806-20 is about $\sim(1.8-4.6)\times10^{46}$ erg \citep{2005Natur.434.1098H}, which accounts for $\sim(0.09-0.92)\%$ of the total magnetic energy. 

In our model, the energy release during the state transition process (or catastrophic escape) of the flux rope is briefly calculated by the energy difference value between the critical point and the stable point (or infinity). For the state transition case shown in Figure 1, the energy difference between the two typical states, the critical state (c) and the stable state (f), could represent the total amount of released magnetic energy. In the state transition process, as explicitly introduced in Paper II, the energy release fraction is expressed by $\Delta W_t/W_t(h_c) = \left[W_t(h_c)-W_t(h_f)\right]/W_t(h_c)$. The calculations under the circumstance of Figure 1 show that the percentage of magnetic energy release is $\Delta W_t/W_t (h_c) \sim 0.42\%$. 

For the catastrophic escape case in Figure 3, the corresponding energy release fraction could be obtained by the difference value between the critical state (c) and a hypothesized state at infinity, which is expressed as $\Delta W_t/W_t(h_c)=\left[W_t(h_c)-W_t( \infty)\right]/W_t(h_c)$. The calculated magnetic energy release percentage in this case is $\Delta W_t/W_t(h_c) \sim 0.51\%$. However, the actual total energy release from magnetosphere during these two processes could be larger than the percentages above. The magnetic reconnection would take place along the current sheet and could release extra energy, which is not taken into consideration in our model. It is likely that under the two different circumstances in our model, the energy release is enough to supply for the radiation of the most extreme giant flare from SGR 1806-20, which accounts for $\sim(0.09-0.92)\%$ of the total magnetic energy possessed in the magnetar.

\section{CONCLUSION}

In order to reveal the influence to giant flares by the variations of multipolar structure at the magnetar surface, we have investigated the mechanism for an eruption, which is demonstrated by a catastrophic loss of equilibrium of a helically twisted flux rope embedded in a force-free magnetosphere. We finally find that the variation of the multipolar magnetic field geometry on the magnetar surface could lead to the catastrophic rearrangement of the magnetosphere and result in giant flare events. A multipolar magnetic field configuration is taken into account as one of the boundary conditions, which includes the contribution from both a dipolar component and a high order multipolar component. The flux rope gradually evolves in quasi-static equilibrium states, corresponding to the variations at the neutron star surface. Once the flux rope reaches the critical point where the equilibrium cannot be maintained, a current sheet begins to appear; and the flux rope is out of balance and erupts along the current sheet catastrophically. The originally closed magnetic fields might then be unfolded (especially in the escape case), and the magnetic energy stored in the magnetosphere would be released instantaneously.

The observations on the 2004 giant flare of SGR 1806-20 showed the highest energy release. The total magentic energy estimated in the magnetosphere of this magnetar is $\sim(5-20)\times 10^{48}$ erg, and the energy released during the giant flare is at the percentage of $\sim(0.09-0.92)\%$. We have investigated the variation of the multipolar structure at the magnetar surface in our model, then provide the energy acculation mechanism before eruption, and finally demonstrate certain cases that could well explain the enormous energy release observed in giant flares. The state transition case shows a $\sim 0.42\%$ energy release of the total magnetic energy, which is in the range of the estimated energy release percentage. Further more, the catastrophic escape case obviously could provide higher energy release percentage ($\sim$0.51$\%$ as reported), which is likely to account for the observed giant flares from magnetars.

In addition, another important factor, the magnetic reconnection, is likely to take place along the assumed undisturbed current sheet. The influence caused by the magnetic reconnection could be demonstrated through an additional time-dependent calculation in our flux rope model, which brings in another important physical quantity, the Mach number of the plasma (expressed as $MA$). As a forecast of the magnetic reconnection situation, in the multipolar magnetic field boundary conditions we adopted, the energy release percentage could be 1$\%$ when $MA=0.005$, and could reach to 2$\%$ when $MA=0.01$. Magnetic reconnection will definitely raise the proportion of energy release in the cases we mentioned above, on which we will submit a detailed report subsequently \citep{2018Huang...} \citetext{in prep.}.

We are grateful to the anonymous referee's thoughtful comments, which facilitate great improvement to this paper. This work has been supported in part by the National Natural Science Foundation of China (grants 11173046, 11590784, 11373064, 11521303, 11203055, 11773054and 11733010), Yunnan Natural Science Foundation (grant 2014HB048), Yunnan Province (2017HC018) and Key Research Program of Frontier Sciences, CAS (Grant No. QYZDJ-SSW-SLH057).

\label{lastpage}

\bibliographystyle{apj}
\bibliography{ref}

\begin{thebibliography}{}
\expandafter\ifx\csname natexlab\endcsname\relax\def\natexlab#1{#1}\fi

\bibitem[{{Duncan} \& {Thompson}(1992)}]{1992ApJ...392L...9D}
{Duncan}, R.~C., \& {Thompson}, C. 1992, \apjl, 392, L9

\bibitem[{{Feroci} {et~al.}(2001){Feroci}, {Hurley}, {Duncan}, \&
  {Thompson}}]{2001ApJ...549.1021F}
{Feroci}, M., {Hurley}, K., {Duncan}, R.~C., \& {Thompson}, C. 2001, \apj, 549,
  1021

\bibitem[{{Forbes}(2010)}]{2010hssr.book..159F}
{Forbes}, T. 2010, {Models of coronal mass ejections and flares}, ed. C.~J.
  {Schrijver} \& G.~L. {Siscoe} (Cambridge University Press), 159

\bibitem[{{Gavriil} {et~al.}(2002){Gavriil}, {Kaspi}, \&
  {Woods}}]{2002Natur.419..142G}
{Gavriil}, F.~P., {Kaspi}, V.~M., \& {Woods}, P.~M. 2002, \nat, 419, 142

\bibitem[{{Gill} \& {Heyl}(2010)}]{2010MNRAS.407.1926G}
{Gill}, R., \& {Heyl}, J.~S. 2010, \mnras, 407, 1926

\bibitem[{{Huang} \& {Yu}(2014{\natexlab{a}})}]{2014ApJ...787..175H}
{Huang}, L., \& {Yu}, C. 2014{\natexlab{a}}, \apj, 787, 175

\bibitem[{{Huang} \& {Yu}(2014{\natexlab{b}})}]{2014ApJ...796....3H}
---. 2014{\natexlab{b}}, \apj, 796, 3

\bibitem[{{Huang} {et~al.}(2018){Huang}, {Yu}, {Yao}, \& {Shen}}]{2018Huang...}
{Huang}, L., {Yu}, C., {Yao}, G.~R., \& {Shen}, Z.~Q. 2018, (in preparation)

\bibitem[{{Hurley} {et~al.}(2005){Hurley}, {Boggs}, {Smith}, {Duncan}, {Lin},
  {Zoglauer}, {Krucker}, {Hurford}, {Hudson}, {Wigger}, {Hajdas}, {Thompson},
  {Mitrofanov}, {Sanin}, {Boynton}, {Fellows}, {von Kienlin}, {Lichti}, {Rau},
  \& {Cline}}]{2005Natur.434.1098H}
{Hurley}, K., {Boggs}, S.~E., {Smith}, D.~M., {et~al.} 2005, \nat, 434, 1098

\bibitem[{{Kouveliotou} {et~al.}(1999){Kouveliotou}, {Strohmayer}, {Hurley},
  {van Paradijs}, {Finger}, {Dieters}, {Woods}, {Thompson}, \&
  {Duncan}}]{1999ApJ...510L.115K}
{Kouveliotou}, C., {Strohmayer}, T., {Hurley}, K., {et~al.} 1999, \apjl, 510,
  L115

\bibitem[{{Lundquist}(1950)}]{1950Ark...2..361}
{Lundquist}, S. 1950, in Ark. Fys., Vol.~2, 361

\bibitem[{{Lyutikov}(2006)}]{2006MNRAS.367.1594L}
{Lyutikov}, M. 2006, \mnras, 367, 1594

\bibitem[{{Mazets} {et~al.}(1982){Mazets}, {Golenetskii}, {Gurian}, \&
  {Ilinskii}}]{1982Ap&SS..84..173M}
{Mazets}, E.~P., {Golenetskii}, S.~V., {Gurian}, I.~A., \& {Ilinskii}, V.~N.
  1982, \apss, 84, 173

\bibitem[{{Mazets} {et~al.}(1979){Mazets}, {Golentskii}, {Ilinskii}, {Aptekar},
  \& {Guryan}}]{1979Natur.282..587M}
{Mazets}, E.~P., {Golentskii}, S.~V., {Ilinskii}, V.~N., {Aptekar}, R.~L., \&
  {Guryan}, I.~A. 1979, \nat, 282, 587

\bibitem[{{Mereghetti}(2008)}]{2008A&ARv..15..225M}
{Mereghetti}, S. 2008, \aapr, 15, 225

\bibitem[{{Mereghetti} {et~al.}(2005){Mereghetti}, {G{\"o}tz}, {von Kienlin},
  {Rau}, {Lichti}, {Weidenspointner}, \& {Jean}}]{2005ApJ...624L.105M}
{Mereghetti}, S., {G{\"o}tz}, D., {von Kienlin}, A., {et~al.} 2005, \apjl, 624,
  L105

\bibitem[{{Palmer} {et~al.}(2005){Palmer}, {Barthelmy}, {Gehrels}, {Kippen},
  {Cayton}, {Kouveliotou}, {Eichler}, {Wijers}, {Woods}, {Granot}, {Lyubarsky},
  {Ramirez-Ruiz}, {Barbier}, {Chester}, {Cummings}, {Fenimore}, {Finger},
  {Gaensler}, {Hullinger}, {Krimm}, {Markwardt}, {Nousek}, {Parsons}, {Patel},
  {Sakamoto}, {Sato}, {Suzuki}, \& {Tueller}}]{2005Natur.434.1107P}
{Palmer}, D.~M., {Barthelmy}, S., {Gehrels}, N., {et~al.} 2005, \nat, 434, 1107

\bibitem[{{Press} {et~al.}(2007){Press}, {Teukolsky}, {Vetterling}, \&
  {Flannery}}]{2007Pre...}
{Press}, W.~H., {Teukolsky}, S.~A., {Vetterling}, W.~T., \& {Flannery}, B.~P.
  2007, in {Numerical Recipes.}, {Third} edn. ({Cambridge: Cambridge Univ.
  Press})

\bibitem[{{Priest} \& {Forbes}(2000)}]{2000Pri...}
{Priest}, E., \& {Forbes}, T. 2000, in {Magnetic Reconnection. MHD Theory and
  Applications.} ({Cambridge: Cambridge Univ. Press})

\bibitem[{{Terasawa} {et~al.}(2005){Terasawa}, {Tanaka}, {Takei}, {Kawai},
  {Yoshida}, {Nomoto}, {Yoshikawa}, {Saito}, {Kasaba}, {Takashima}, {Mukai},
  {Noda}, {Murakami}, {Watanabe}, {Muraki}, {Yokoyama}, \&
  {Hoshino}}]{2005Natur.434.1110T}
{Terasawa}, T., {Tanaka}, Y.~T., {Takei}, Y., {et~al.} 2005, \nat, 434, 1110

\bibitem[{{Thompson} \& {Duncan}(1995)}]{1995MNRAS.275..255T}
{Thompson}, C., \& {Duncan}, R.~C. 1995, \mnras, 275, 255

\bibitem[{{Thompson} \& {Duncan}(1996)}]{1996ApJ...473..322T}
---. 1996, \apj, 473, 322

\bibitem[{{Thompson} {et~al.}(2002){Thompson}, {Lyutikov}, \&
  {Kulkarni}}]{2002ApJ...574..332T}
{Thompson}, C., {Lyutikov}, M., \& {Kulkarni}, S.~R. 2002, \apj, 574, 332

\bibitem[{{Woods} \& {Thompson}(2006)}]{2006csxs.book..547W}
{Woods}, P.~M., \& {Thompson}, C. 2006, {Soft gamma repeaters and anomalous
  X-ray pulsars: magnetar candidates}, ed. W.~H.~G. {Lewin} \& M.~{van der
  Klis}, 547--586

\bibitem[{{Xing} \& {Yu}(2011)}]{2011ApJ...729....1X}
{Xing}, Y., \& {Yu}, W. 2011, \apj, 729, 1

\bibitem[{{Yu}(2011)}]{2011MNRAS.411.2461Y}
{Yu}, C. 2011, \mnras, 411, 2461

\bibitem[{{Yu}(2012)}]{2012ApJ...757...67Y}
---. 2012, \apj, 757, 67

\bibitem[{{Yu} \& {Huang}(2013)}]{2013ApJ...771L..46Y}
{Yu}, C., \& {Huang}, L. 2013, \apjl, 771, L46

\end{thebibliography}
\end{CJK}
\end{document}